\documentclass[reprint, amsmath, amssymb, aps, prb]{revtex4-1}

\usepackage[]{graphicx}      
\usepackage{bm}             	
\usepackage{times}			
\usepackage{tabularx}
\usepackage{color}
\usepackage{dcolumn}		
\usepackage{amsmath}
\usepackage{amssymb}
\usepackage{amsfonts}
\usepackage{times}
\usepackage{tabularx}
\usepackage{xcolor,colortbl}
\newcommand{\IEF}{Institut d'Electronique Fondamentale, CNRS, Univ. Paris-Sud, Universit\'e Paris-Saclay, 91405 Orsay, France}

\begin{document}
\title{Ferromagnetic resonance of exchange-coupled perpendicularly magnetized bilayers}

\author{Thibaut Devolder}
\email{thibaut.devolder@u-psud.fr}
\affiliation{\IEF}

\date{\today}                                           
%
%
\begin{abstract}
Strong ferromagnetic interlayer exchange couplings $J$ in perpendicularly magnetized systems are becoming increasingly desirable for applications. We study whether ferromagnetic interlayer exchange couplings can be measured by a combination of broadband ferromagnetic resonance methods and magnetometry hysteresis loops. For this we model the switching and the eigenexcitations in bilayer systems comprising a soft layer coupled to a thicker harder layer that possesses higher perpendicular magnetic anisotropy. For large $J >0$ the switching fields are essentially independent of $J$ but the frequency of the optical eigenmode of the bilayer and the linewidth of the acoustical and optical eigenmode are directly sensitive to the coupling. We derive a corpus of compact analytical expressions to analyze these frequencies, their linewidth and discuss the meaning thereof. We illustrate this corpus on a system mimicking the fixed layers of a magnetic tunnel junction meant for spin torque applications.
\end{abstract}

\keywords{ferromagnetic resonance, interlayer exchange coupling, magnetic tunnel junction, perpendicular anisotropy}

\maketitle

%
%

\section{Introduction}

The next generations of spin-torque operated magnetic memory cells will rely on perpendicular magnetic anisotropy (PMA) magnetic tunnel junctions (MTJ) \cite{ikeda_perpendicular-anisotropy_2010}, because of the superior scalability of this configuration \cite{khvalkovskiy_basic_2013}. The correct sensing of the stored information requires reference layers with a resilient magnetization orientation that should be insensitive to thermal fluctuations, magnetic fields and spin-torques. In practice this translates into reference layer requirements for a sufficient anisotropy, a sufficient damping and no stray field. In addition, the layout of the reference system must ensure a high tunnel magneto-resistance (TMR). 
As a result, optimized MTJs \cite{sun_effect_2011, gan_perpendicular_2014, worledge_spin_2011, jung_interlayer_2012, devolder_performance_2013, moriyama_tunnel_2010, you_co/ni_2012, ishikawa_magnetic_2013, ishikawa_co/pt_2014, swerts_beol_2015, devolder_time-resolved_2016} rely on composite reference systems, where each of the previously mentioned feature is optimized by some sub-system. The TMR is generally optimized by using an FeCoB\cite{kanai_magnetic_2014} spin polarizing layer (layer 1 of thickness $t_1$), while the high anisotropy is usually provided by an fcc (111) cobalt-based multilayer (layer 2 of thickness $t_2$). Because these two layers have different crystal structures a spacer layer is needed. 

However the role of the spacer layer is also to promote a high interlayer exchange coupling $J$  between its adjacent layers 1 and 2. Indeed, the sole criterion of maximal TMR would argue \cite{ishikawa_co/pt_2014, wisniowski_effect_2008} for a large $t_1$, leading to the loss of perpendicular magnetization at remanence when the demagnetizing energy $\mu_{0}M_{S}^2 /2$ overcomes the interface anisotropy energy $K_S / t_1$. This can be relaxed and one can use thicker spin polarizing layer if one complements its interface anisotropy by a ferromagnetic interlayer exchange coupling $J/t_1$ with the high anisotropy layer. As a result, the measurement of large and ferromagnetic interlayer exchange coupling in soft layer / spacer / hard layer composites is of importance.
Unfortunately, conventional magnetometry methods are ineffective to measure large positive $J$: strong couplings force the layers 1 and 2 to switch in synchrony in a rigid manner, such that the loops are not informative with respect to the amplitude of the coupling. Besides, the coercivities of real systems are often extrinsic  and largely influenced by the defects, in addition to being affected by thermal activation; as a result the switching fields in hysteresis loops can only be used to provide qualitative information on the anisotropies and the coupling.

In this paper, we study whether the ferromagnetic resonance modes of the composite can be used to quantify the coupling by either looking at the eigenmode frequencies or their linewidth in addition to the hysteresis loops. We first solve the system exactly for material properties that are typical in order to classify the possible class of behaviors. We then derive approximate analytical formulas meant to ease the analysis of future experimental studies of the eigenmode frequencies and their linewidth. 

We model samples consisting of a soft layer / spacer / hard layer sequence with perpendicular anisotropies. We describe the system as a set of two coupled macrospin labelled $i=1,~ 2$, with thicknesses $t_i$, magnetizations $M_{Si}$, damping parameters $\alpha_i$, magnetocrystalline anisotropy fields $H_{ki}$, normalized magnetization components $\{m_{xi},~m_{yi}, ~m_{zi}\}$ and coupled through a bilinear interlayer exchange energy $J$. $J$ is supposed to be much smaller than the intralayer exchange stiffness so that the macrospin approximations remain valid within each layer \cite{pashaev_ferromagnetic-resonance_1991}. For simplicity we will often gather the anisotropy and the demagnetizing energies together, by writing $H_{ki}^{\textrm{eff}}={H_{ki}} - {M_{Si}}$ for each layer.
The layer $i=1$ is chosen as the softest and the thinnest layer. The areal energy of the system (in units of $\mu_0$) is the sum of the anisotropy and demagnetizing energies :
$({1}/{2}) \left(-H_{k1}^{\textrm{eff}} {M_{S1}} {m_{z1}}^2 {t_{1}}-H_{k2}^{\textrm{eff}} {M_{S2}}
   {m_{z2}}^2 {t_{2}}\right)$
and of the Zeeman energy $ -{H_{z}} ({M_{s1}} {m_{z1}} {t_{1}}+   {M_{s2}} {m_{z2}} {t_{2}})$  and  the interlayer exchange coupling energy $-J_0 ({m_{x1}} {m_{x2}}+{m_{y1}}{m_{y2}}+{m_{z1}} {m_{z2}})$
where we have written $J_0 = J / \mu_0 $, with $J$ in $\textrm{J/m}^2$ and all other terms also expressed in SI units. 
 The configuration of the system is found by minimizing the total energy, while its eigenexcitations are found by linearizing the magnetization dynamics equation about that energy minimum. The analytical expressions for the frequencies and the critical fields are derived for vanishing damping parameters. 

\begin{figure}
\includegraphics[width=9 cm]{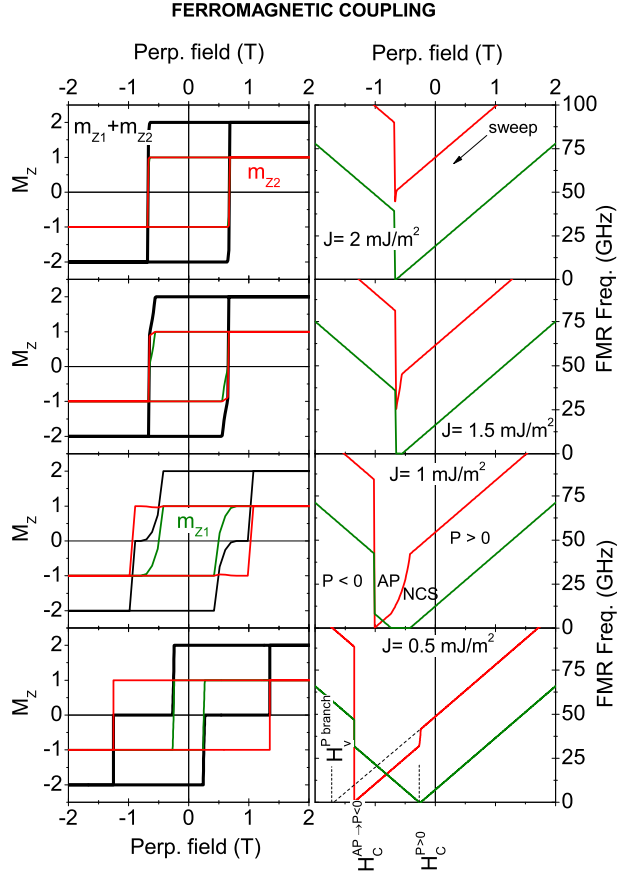}
\caption{Hysteresis loops (left panels, back and forth field sweeps) and eigenexcitations (right panels, decreasing field only) of a soft/hard PMA composite bilayer for ferromagnetic interlayer exchange coupling. $P$, $AP$ and $NCS$  stand for parallel, antiparallel and non collinear states. The black loops are the sum of the normalized magnetizations of the two layers.}
\label{allFERRO}
\end{figure}

\section{Classification of the possible behaviors}

Let us first look at some typical behaviors by solving the system numerically on a given set of material parameters. The calculations are done with parameters mimicking an FeCoB layer as soft layer (i.e. layer 1) and $[$Co/Pt$]_N$ multilayer as hard layer (i.e. layer 2). In the numerical calculations, the chosen parameters are $M_{S1}=10^6$ A/m, $M_{S2}=8\times 10^5$ A/m, $H_{k2}=2\times 10^6$ A/m, $t_1=2$ nm, $t_2=5$ nm. The interlayer exchange coupling was varied from strongly antiferromagnetic ($J=-1.5$ mJ/m$^2$) to strongly ferromagnetic ($J=2$ mJ/m$^2$). The field is swept from positive (favoring a so-called Parallel "$\textrm P>0$" state) to negative (favoring a so-called reversed Parallel "$\textrm P<0$" state). 
In all figures except Fig.~\ref{figIMA}, the magneto-cristalline anisotropy of layer 1 is $H_{k1}=10^6$ A/m, leading to a zero effective anisotropy. In Fig.~\ref{figIMA}, we shall use $H_{k1} = 6\times10^5$ A/m (implying \textit{negative} effective anisotropy) in order to describe also the systems in which the perpendicular remanence of the layer 1 is obtained by the coupling and not by its sole anisotropy.

\subsection{Case of ferromagnetic interlayer coupling}
The case of ferromagnetic interlayer exchange coupling is displayed in Fig.~\ref{allFERRO}. The common features for $J>0$ are that the coercivities have always the normal sign (positive convention) and that the switching of a given layer is always accompanied with an increase of the frequency of its eigenexcitation. Let us look at the details. \\
In case of very strong coupling (Fig.~\ref{allFERRO}, top panels), the soft and hard layers switch simultaneously and always keep a magnetization fully perpendicular to the sample plane: there is a direct transition from the $\textrm P>0$ to the $\textrm P>0$ state. The lowest frequency eigenmode (the acoustical excitation) softens to zero at this unique (positive) switching field $H_{C,~1}=H_{C,~2}$. \\
When the coupling is reduced to 1.5 mJ/m$^2$, the two layers do not switch in synchrony any longer: the soft layer first tilts its magnetization creating a Non Collinear State (NCS), and it saturates only when hard layer switches also; there is a tiny field interval during which the magnetizations are no longer collinear to each other, and the acoustical mode stays soft (i.e. $\omega=0$) during that interval. \\
When the coupling is further reduced to 1 mJ/m$^2$, the two layers start to switch at distinct fields. NCS magnetizations states tilted from the perpendicular axis are possible for the two layers. An AntiParallel (AP) configuration occurs in part of the field interval between the two coercivities. The highest frequency eigenmode (the optical excitation) evolves non linearly with the applied field in that interval. One of the AP eigenmode softens at the corresponding AP to $\textrm P<0$ switching field.\\ 
Finally when the coupling is reduced to a weak value of 0.5 mJ/m$^2$, the two layers switch independently. Only collinear states are possible, but a large interval allows for an antiparallel (AP) situation. The $\textrm P>0$ to AP switching of the soft layer induces a step-like reduction of the eigenexcitation frequency of the optical mode, whose amplitude is mostly localized in the hard layer.

\begin{figure}
\includegraphics[width=9 cm]{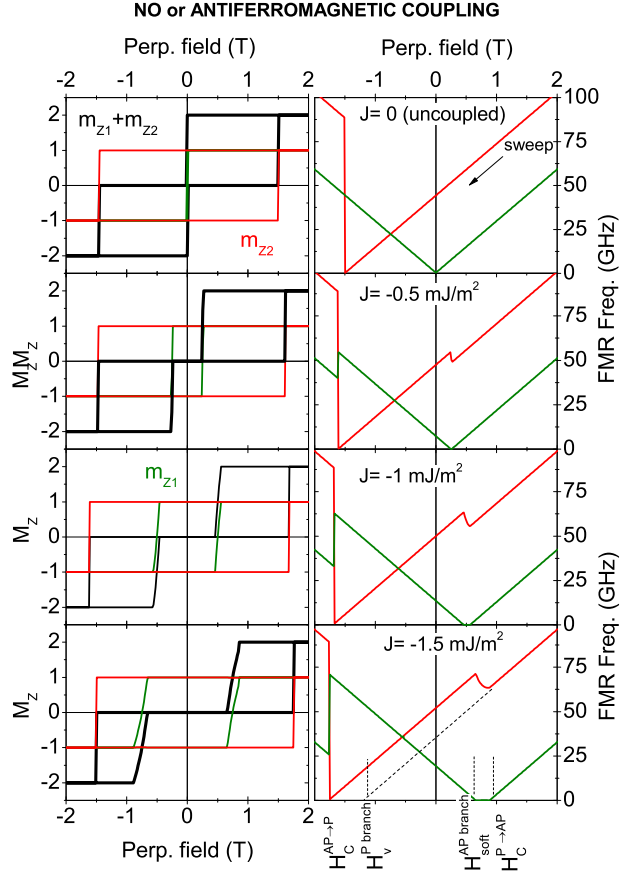}
\caption{Hysteresis loops (left panels, back and forth field sweeps) and eigenexcitations (right panels, decreasing field only) of a soft/hard PMA composite bilayer for uncoupled films and antiferromagnetically coupled films.}
\label{allANTIFERRO}
\end{figure}

\subsection{Case of antiferromagnetic interlayer coupling}
The cases of zero and antiferromagnetic interlayer exchange couplings are displayed in Fig.~\ref{allANTIFERRO}. The soft layer coercivities is decreased and its switching is always accompanied with an increase of the frequency of the eigenexcitation of the hard layer. Let us look at the details. \\
When the layers are not coupled (top panels), they switch separately at their respective effective anisotropy fields, and their corresponding ferromagnetic resonance modes are independent. Parallel and antiparallel collinear configurations both occur during a loop, but the magnetizations always stay perpendicular to the plane. \\
When the antiferromagnetic coupling is turned on to a weak value of -0.5 mJ/m$^2$, the soft layer coercivity decreases and can change sign (switching happens at a \textit{positive} before reaching zero field). The soft layer $\textrm P>0$ to AP switching  increases the frequency of the hard layer eigenexcitation. The amplitude of this frequency jump increases with the strength of $|J|$. \\
When the antiferromagnetic coupling is further increased to -1 and -1.5 mJ/m$^2$, non collinear states (NCS) become possible again. There is a gradual increase of the $\textrm P>0$ to NCS coercivity of the soft layer, and of the frequency impact of a layer's switching on the other. \\
Finally for very strong antiferromagnetic coupling (bottom panel), NCS occur at the onset of the soft layer switching; this comes with a rounding of the corresponding frequency jumps. The AP state occurs in a large field interval.

In summary, the specific example calculated above illustrates most of the possible behaviors.
Starting from $H_z>0$, there are 4 possible states $\textrm P>0$, $\textrm P<0$, AP and non collinear configurations NCS. The corresponding state diagram is displayed in Fig.~\ref{criticalFIELDScalc}. The boundaries of this state diagram depends on the material properties. Noticeably, the NCS states occur more frequently when the layers' easy axes are different, as illustrated in Fig.~\ref{figIMA} in which the layer 1 has been chosen with an easy plane (i.e. $H_{k1}^\textrm{eff} <0$). Let us derive the boundaries of the state diagrams in an analytical manner.

\section{Analytical model in the parallel states} \label{analytical}
From the previous section, we have seen that states with parallel magnetizations are present in large field intervals, especially in the case of ferromagnetic coupling. With the objective of easing the analysis of experimental data, we spend the next sections to derive a corpus of analytical expressions that describe the switching fields and the eigenexcitation frequencies in the $\textrm P>0$ state (this section) and AP (next section) case. The first step is to linearize the magnetization dynamics equation about that configuration when it is the energy minimum, and then take the Hessian matrix of the total energy of the system to find how the effective fields depend on the dynamic magnetization. In the parallel state, the eigenfrequencies $\Re(\omega)$, the half linewidth $-\Im(\omega)$ and the eigenmodes of the soft/hard composite are the complex eigenvalues $\omega = \Re(\omega) + i \Im(\omega)$ and the eigenvectors of the so-defined dynamical matrix:

$$
\left(
\begin{array}{cccccc}
 \alpha_1 \tilde{H}_1 & \tilde{H}_1 & 0 & \frac{\alpha_1 J_0}{{M_{S1}} {t_1}} &
   -\frac{J_0}{{M_{S1}} {t_1}} & 0 \\
 - \tilde{H}_1 & \alpha_1 \tilde{H}_1 & 0 & \frac{J_0}{{M_{S1}}
   {t_1}} & \frac{\alpha_1 J_0}{{M_{S1}} {t_1}} & 0 \\
 0 & 0 & 0 & 0 & 0 & 0 \\
\frac{\alpha_2 J_0}{{M_{S2}} {t_2}} & -\frac{J_0}{{M_{S2}} {t_2}} & 0 & \alpha_2 \tilde{H}_2 & \tilde{H}_2 & 0 \\
 \frac{J_0}{{M_{S2}} {t_2}} & \frac{\alpha_2 J_0}{{M_{S2}} {t_2}} & 0 &  - \tilde{H}_2 & \alpha_2 \tilde{H}_2 & 0 \\
 0 & 0 & 0 & 0 & 0 & 0 \\
\end{array}
\right)
$$

where we have defined $\tilde{H}_i =  H_{ki}^{\textrm{eff}} +H_z +\frac{J_0}{{M_{Si}} {t_i}}$. Note that as our physical problems has only 4 independent variables which are the orientation angles of the two magnetizations, the above $6\times6$ matrix is only of rank 4:  two of its eigenvalues are null and the eigenvalues are two-by-two degenerate. Writing the problem in circular coordinates is unpractical because the ground states are close to the $\theta_i=0$ and $\theta=\pi$ positions where circular coordinates are singular.
Our notation is such that $ \alpha_i = - \frac{\Im(\omega_i)}{Re(\omega_i)} $ for perpendicularly magnetized states in the absence of coupling (i.e. $J=0$).
From the antisymmetric character of the four $3 \times 3$ sub-quarters of the dynamical matrix, it results that the eigenmodes will have no ellipticity in this $P>0$ state, and that consequently the frequency versus field curves will have a slope being the gyromagnetic ratio $\gamma_0$. The knowledge of the critical fields $H_\textrm{crit}$ at which these modes soften is thus enough to describe the eigenmode frequencies which will then follow $\omega = \gamma_0 (H_z - H_\textrm{crit})$. 

In the specific case of no coupling, the dynamical matrix yields the usual ferromagnetic resonance frequencies for the two layers : $\omega_{\textrm{FMR}, ~J=0} = \gamma_0 (H_{ki}^{\textrm{eff}} + H_z)$ which gives (negative) switching fields of $H_{Ci}=-H_{ki}^{\textrm{eff}}$ when these modes soften.  \\
The situation is more complex when there is a finite coupling.
Let us focus on the case of the $\textrm{P}>0$ ground state, and first find the softening field at which this state looses stability. They are the fields at which at least one eigenexcitation of the system softens to zero frequency. 

\subsection {$\textrm P>0$ parallel state softening fields}
\subsubsection{Identical layers}s

In the case of identical layers (i.e. $M_{S1}=M_{S2}=M_S$, $t_1=t_2$ and $H_{k1}=H_{k2}=H_{k}$), there are two softening fields at which the hypothetical $\textrm{P}>0$ state looses stability. The first one is simply: 
$$H_{\textrm C, ~1=2}^{\textrm P>0} = -H_k^{\textrm{eff}}$$  
The stability analysis indicates that this is a real switching field for ferromagnetic coupling, and it induces a transition to $\textrm P<0$, AP or NCS. Note that this switching field is independent of $J$, which confirms that conventional magnetometry is inoperative to quantify $J$; Qualitatively, the switching field is independent of $J$ because if we virtually "cut" a layer in two halves, this does not change the physics as long as the two parts are ferromagnetically coupled.

The second softening field is 
\begin{equation}
H_{v, ~1=2}^{\textrm{P~branch}}=H_{\textrm C, ~1=2}^{\textrm P>0} - {\frac{2 J_0}{M_S t}}  
\label{symSOFTfield}
\end{equation}
The subscript $v$ stands for \textit{virtual}; indeed generally $H_{v, ~1=2}^{\textrm{P~branch}}$  is not a switching field since the $\textrm{P}>0$ state has lost stability (hence has disappeared) before this field is actually reached during a field sweep. It is however an interesting quantity as it is the zero frequency extrapolation of the optical excitation of the symmetric bilayer in the $\textrm{P}>0$ branch (i.e. this is the analogous of the Perpendicular Standing Spin Wave in single uniform films). \\

\subsubsection{Non identical layers}
In practice, we deal with soft and hard layers that can have very different properties. To describe this non-symmetric case, it is useful to define the notations gathered in Table~\ref{notations}.\\

\begin{table}[ht]
\caption{Notations used when describing non identical layers.} 
\centering 
\begin{tabular}{| c | c |  } 
\hline
Physical property  & Formula  \\ [0.5ex] 
\hline 
 & \\ [0.1ex]
Miscompensation of moments & $\Delta Mt = M_{S1} t_1 - M_{S2} t_2$  \\ 
Difference in anisotropies & $\Delta H_k^{\textrm{eff}} = H_{k1}^{\textrm{eff}} - H_{k2}^{\textrm{eff}}$   \\ [1ex] \hline 
 & \\ [0.1ex]
Total moment & $\Sigma Mt = M_{S1} t_1 + M_{S2} t_2$   \\
Total anisotropy & $\Sigma H_k^{\textrm{eff}} = H_{k1}^{\textrm{eff}} + H_{k2}^{\textrm{eff}}$   \\
"average" square moment & ${M_S}^2 t^2 = M_{S1} M_{S2} t_1 t_2$  \\ [1ex] 
\hline 

\end{tabular}
\label{notations} 
\end{table}

The expressions of the eigenmode frequencies are $\gamma_0 (H_z-H_{\textrm{soft}, ~1\neq 2}^{\textrm P>0})$, and they soften at critical fields that can be expressed as:
\begin{eqnarray}\nonumber
H_{\textrm{soft}, ~1\neq 2}^{\textrm P>0}  & =  - \frac{\Sigma H_k^{\textrm{eff}}}{2}-\frac{{J_0\Sigma
   Mt}}{2 {Ms}^2 t^2} \\
  &  \pm \frac{\sqrt{J_0^2 
   {\Sigma Mt}^2-2 {\Delta H_k} {\Delta Mt} J_0 {Ms}^2
   t^2+{\Delta H_k}^2 {Ms}^4 t^4}}{2 {Ms}^2 t^2}
   \label{Hsoft_1}
   \end{eqnarray}

\begin{figure}
\includegraphics[width=9cm]{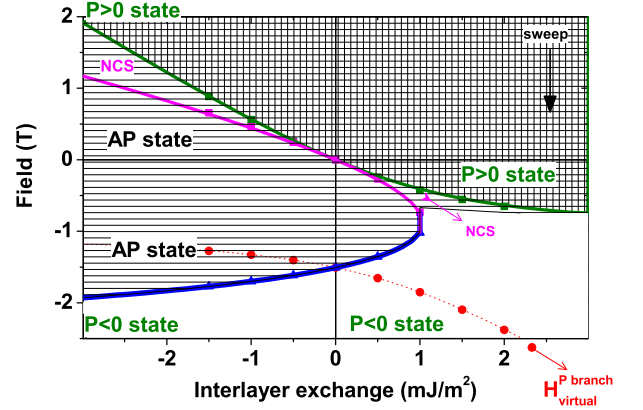}
\caption{State diagram derived from the numerical simulations (symbols) and from the analytical models (lines). The out-of-plane field is swept down after a start at $H_z>0$. The notations $\textrm P>0$, $\textrm P<0$ $\textrm{AP}$ and $\textrm{NCS}$ stand for parallel with positive full remanence, parallel with negative full remanence, antiparallel and non collinear states. The critical fields are Eq.~\ref{Hsoft_1} (green and red lines) and Eq.~\ref{Hsoft_AP} (blue and magenta lines). These lines merge in the point defined by Eq.~\ref{Jmax_for_AP}. }
\label{criticalFIELDScalc}
\end{figure}

Note that once again only one of these two fields is a real switching field: it is the first one that is reached during a field sweep. The other softening field is anyway an interesting quantity. Indeed  it is the zero frequency extrapolation of the optical excitation of the bilayer. These two fields are displayed in red and green in the Figs.~\ref{criticalFIELDScalc} and \ref{figIMA}. 
Two limits of the previous equations are worth mentioning.
\paragraph{Non identical layers, limit of large coupling.}

In the limit of large ferromagnetic coupling (i.e. $J_0^2   {\Sigma Mt}^2$ much greater than the other terms under the square root of eq.~\ref{Hsoft_1} and $J>0$), the first softening field is a switching field away from $\textrm P>0$ to either $\textrm AP$, NCS or $\textrm P<0$), which reduces to:
\begin{equation}
H_{\textrm C,~1\neq2~J>>1} \approx -\frac{\Sigma H_k^{\textrm {eff}}}{2} -\frac{{\Delta H_k} {\Delta Mt}}{2 {\Sigma Mt}} +O(\frac{1}{J_0})
\label{asymSOFTfield1largeJ}
\end{equation}

The expression of $H_{\textrm C,~1\neq2~J>>1}$ is only an asymptotic limit [see Fig.~\ref{figdamping}(a)] which requires $J >> 2~\textrm{mJ/m}^2$, i. e. direct layer-to-layer exchange coupling, or pinhole mediated exchange coupling. In that case, the first term of the right hand side of the above expression means that the switching field is essentially the average of these of the two layers when uncoupled, and the second term is a correction that scale with the asymmetry of their properties. This recalls the results for in-plane magnetized systems \cite{heinrich_structural_1988} in which a so-called bilayer scaling parameter can be used to describe how the bilayer cocercivity can be calculated from an average of the properties of the two layers. \\
In this asymptotic limit of strong ferromagnetic coupling, the second softening field reduces to a virtual field that is:
\begin{equation}
H_{v,~1\neq2~J>>1}^\textrm{P~branch} \approx H_{\textrm  C,~1\neq2~J>>1} - \frac {J_0 \Sigma Mt}{{Ms}^2 t^2}~.
\label{asymSOFTfield2largeJ}
\end{equation}
This field corresponds to the zero frequency intercept of the optical (PSSW-like) branch of the bilayer.

\paragraph{Non identical layers, limit of weak coupling.} 
In the limit of weakly exchanged very asymmetric systems (i.e. $J_0^2   {\Sigma Mt}^2$ much smaller than the other terms in the square root Eq.~\ref{Hsoft_1}, in practice for $J << 2~\textrm{mJ/m}^2$), the softening fields reduce to:

\begin{equation}
H_{\textrm C,~1\neq2~J<<1}^{\textrm  P >0} \approx -{H_{k1}^{\textrm {eff}}}-\frac{J_0}{{M_{S1}} {t_1}}
\label{asymSOFTfield1}
\end{equation}

and \begin{equation}
H_{v,~1\neq2~J<<1}^\textrm {P~branch}  \approx -{H_{k2}^{\textrm{eff}}}-\frac{J_0}{{M_{S2}} {t_2}}
\label{asymSOFTfield2}
\end{equation}

These expressions indicate qualitatively that the two eigenmodes are sufficiently distant for little mode hybridation to occur, such that a layer $i$ acts on the layer $j$ as a static exchange biasing field of $\frac{J_0}{{M_{Sj}} {t_j}}$ that needs to be compensated for the stability loss of the $\textrm P>0$ state. When in the $\textrm P>0$ state, a ferromagnetic interlayer exchange coupling is such that each layer stabilizes the other. The coupling simply increases the coercivity of the softest layer (see Eq.~\ref{asymSOFTfield1}).

It is worth noticing the factor of 2 difference in the role of the exchange coupling $J$ in the case of softening fields for strongly asymmetric (Eq.~\ref{asymSOFTfield2}) and symmetric cases (Eq.~\ref{symSOFTfield}). The factor of two recalls that in the symmetric case, the optical excitation involve the motion of the \textit{two} layers with equal amplitude, while in the strongly asymmetric case, each eigenmode is essentially a \textit{one}-layer oscillation under the bias provided by the almost static other layer. This numerical factor than can vary between 1 and 2 depending on the layer-to-layer asymmetry can lead to confusion: indeed it means that the measurement of the sole critical fields (Eq.~\ref{asymSOFTfield1} and \ref{asymSOFTfield2}) can not inform on the value of $J$ unless all other magnetic properties are known.

\section{Analytical model in the case of antiparallel magnetizations}

Similarly, we can derive  the dynamical matrix (not shown) describing the eigenexcitations about the AP position $\{ m_{z1}=-1$, $m_{z2}=1 \}$ where we have assumed that the soft layer switches first. The formalism in this section is only valid when this AP state is visited during a field sweep; note that this is not the case for strong ferromagnetic coupling when only P states occur (Fig.~\ref{criticalFIELDScalc} ).
\subsection{Identical layers}
In the case of identical layers, two softening fields are obtained:
$$H_{\textrm C, 1=2}^{\textrm{AP}}  =\pm \sqrt{
H_{k}^{\textrm{eff}} \times (H_{k}^{\textrm{eff}}-{\frac{2J_0}{M_S t}})}$$
These fields are virtual for strong ferromagnetic coupling (i.e. $ {\frac{2J_0}{Ms t}} > H_{k}^{\textrm{eff}}$) and this AP state is not visited during a field sweep. For small or negative $J$, these two critical fields are real and opposite as intuitively expected for a symmetric bilayer. 

\subsection{Non identical layers}
In the case of non symmetric layers, the two softening fields are:

\begin{eqnarray} 
   \label{Hsoft_AP}
H_{\textrm{soft},~1\neq2}^{\textrm{AP}} & = \frac{\Delta H_k^{\textrm{eff}}}{2} + \frac{J_0\Delta
   Mt}{2 {M_s}^2 t^2}\\
& \pm \frac{\sqrt{J_0^2
   {\Delta Mt}^2-2 {\Sigma H_k^{\textrm{eff}} {\Sigma Mt} J_0 {Ms}^2
   t^2+{\Sigma H_k}^2 {M_s}^4 t^4}}}{2 {M_s}^2 t^2} \nonumber
\end{eqnarray}
where $\pm$ distinguishes the real and the virtual switching fields whose role are interchanged when the sign of $J$ is changed.  The two branches of Eq.~\ref{Hsoft_AP} are plotted in Fig.~\ref{criticalFIELDScalc}. 

\begin{figure}
\includegraphics[width=9cm]{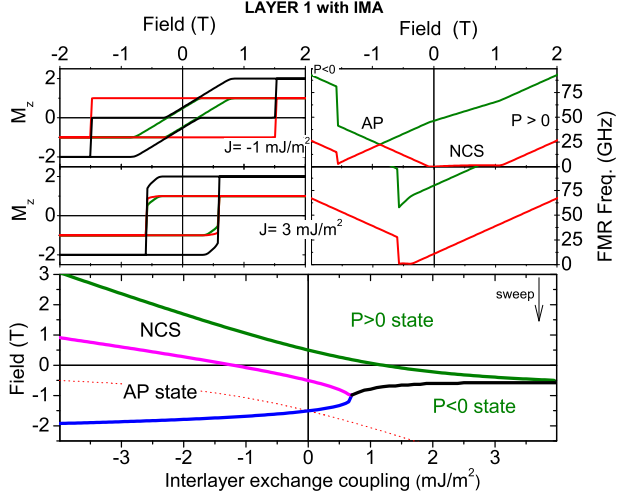}
\caption{Loops and eigeneiexcitation frequencies when the soft layer is decreased and yields a negative effective anisotropy ($H_{k1}= 600$ kA/m and $M_{S1}=$1 MA/m) for two values of the interlayer exchange coupling. Bottom panel: state diagram when the out-of-plane field is swept down after a start at $H_z>0$. The notations $\textrm P>0$, $\textrm P<0$ $\textrm{AP}$ and $\textrm{NCS}$ stand for parallel with positive full remanence, parallel with negative full remanence, antiparallel and non collinear states. The critical fields are Eq.~\ref{Hsoft_1} (green and red lines) and Eq.~\ref{Hsoft_AP} (blue and magenta lines). These lines merge in the point defined by Eq.~\ref{Jmax_for_AP}. The black line is a numerical resolution.}
\label{figIMA}
\end{figure}

It is worth comparing the signs of the first terms of Eq.~\ref{Hsoft_AP} and Eq.~\ref{Hsoft_1}. Indeed the former decreases with $H_{k1}^{\textrm{eff}}$ while the second increases, with the consequence that the NCS pockets in the state phase diagrams (Fig.~\ref{criticalFIELDScalc}) essentially shrink and separate when the layer 1 anisotropy is strengthened, or the two NCS pockets merge in a single one that enlarges when the layer 1 anisotropy is decreased (Fig.~\ref{figIMA}).\\
From the merging of the two branches of Eq.~\ref{Hsoft_AP} it is also interesting to see that the condition for the non existence of an AP state during the hysteresis loop is simply $J_0 \geq J_{\textrm{no~AP}} $ with
\begin{equation}
J_{\textrm{no~AP}} = {M_S}^2 t^2 \Big{[} \frac{\sqrt{M_{S1}t_1}-\sqrt{M_{S2}t_2}}{M_{S1}t_1-M_{S2}t_2} \Big{]}^2 \Sigma H_k^\textrm{eff}
\label{Jmax_for_AP}
\end{equation}
If both layers have positive effective anisotropies, this condition ensures that the only possible states are the P states: in practice this is a condition of proper pinning of the softest layer. It is thus the condition that has to fulfilled in applications where the softest layer ($i=1$) is required to stay magnetized parallel to the hard layer. In our numerical examples, this condition is $\mu_0 J_{\textrm{P~only}} = 1.03~\textrm{mJ/m}^2$.

Two limits of Eq.~\ref{Hsoft_AP} are worth looking at.\\

\paragraph{Non identical layers, limit of large coupling.}

In the limit of large coupling and/or strong asymmetry (i.e. when $J_0^2
   {\Delta Mt}^2$ is much greater than the other terms under the square root in Eq.~\ref{Hsoft_AP}), the softening fields reduce to 

$$H_{\textrm{soft},~1\neq2~J>>1}^{\textrm{AP~branch}} \approx  \delta \frac{J_0\Delta
   Mt}{{M_s}^2 t^2} + \frac{\Delta H_k^{\textrm{eff}}}{2} +  \frac{\Sigma H_k^{\textrm{eff}} \Sigma Mt }{2 \Delta Mt} +O(\frac{1}{J_0})$$
   where $\delta$ is either 0 or 1 for the real and virtual switching fields. The convergence of the expression at large $J$ is slow and one should better use Eq.~\ref{Hsoft_AP} for the asymmetries and the coupling values encountered in practice.

\paragraph{Non identical layers, limit of weak coupling.}
In the limit of weakly exchanged, or nearly symmetric systems, the softening fields reduce to: 
\begin{equation}
H_{\textrm C,~1\neq2~J<<1}^{\textrm{AP}\leftrightarrow \textrm {NCS}} \approx H_{k1}^{\textrm{eff}} - \frac{J_0}{M_{s1} t_1}
\label{HsoftAP1}
\end{equation}
which is the field at which the AP state with $m_{z1}=-1$ looses stability upon increasing field, and
\begin{equation}
H_{v,~1\neq2~J_0<<1}^{\textrm{AP} \rightarrow \textrm{P}<0} \approx - H_{k2}^{\textrm{eff}} + \frac{J_0}{M_{s2} t_2}
\label{HsoftAP2}
\end{equation}
which is the field leading to a switching of the hard layer ($i=2$) towards the $\textrm P <0$ state.
These two above expressions recall the correspond ones in Eq. \ref{asymSOFTfield1} for the instability of the $\textrm P > 0 $ case. They illustrate that for weak $J$ or nearly symmetric systems a minor loop of the soft layer ($i=1$) performed between the $\textrm P > 0 $ and the AP states has an opening of $2H_{k1}^{\textrm{eff}}$ and an offset of $-\frac{J_0}{M_{s1} t_1}$. As it was the case in the P states, such ways of measuring $J$ requires the prior knowledge the layers' magnetic properties to be certain to be in the weakly exchanged nearly symmetric regime.

\section{Effect of exchange coupling on the eigenmode linewidths}
\begin{figure}
\includegraphics[width=9cm]{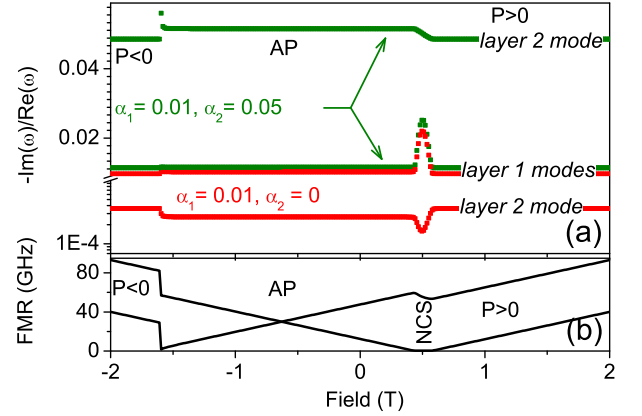}
\caption{Eigenexcitation relative linewidth (a) and frequencies (b) versus field curves for samples whose properties are: $M_{S1}=10^6$ A/m, $M_{S2}=8\times 10^5$ A/m, $H_{k1}=10^6$ A/m, $H_{k2}=2\times 10^6$ A/m, $t_1=2$ nm, $t_2=5$ nm, $\alpha_1=0.01$ and $\alpha_2=0$ (red curves) and 0.05 (green cuves) for an antiferromagnetic coupling of $J=-1$ mJ/m$^2$. When in the NCS state, the eigenmodes have a finite ellipticity, which correlates with a substantial increase to the acoustical mode linewidth.
}
\label{figLinewidth}
\end{figure}

In this last section, we study how the damping of a given layer affects the linewidth of the two eigenmodes thanks to the interlayer exchange coupling $J$. In the numerical evaluations, we use $\alpha_1=0.01$ (mimicking FeCoB \cite{devolder_damping_2013}) and a variable $\alpha_2=0-0.1$ to mimic the effect of the harder layer. As in the isolated layer case, introducing reasonable values of damping does not change the eigenmode frequencies and consequently the critical fields at which the modes soften (Fig.~\ref{figLinewidth}). 

\begin{figure}
\includegraphics[width=9cm]{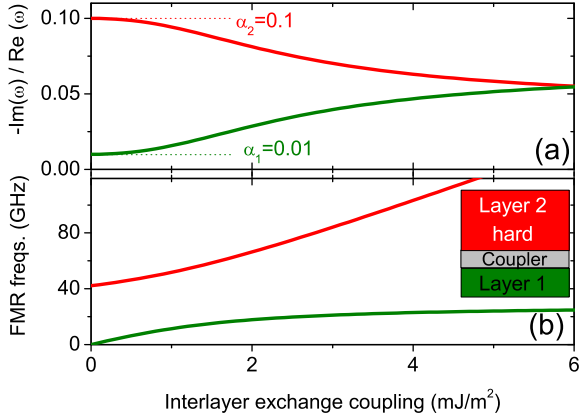}
\caption{Eigenexcitation relative linewidths (a) and frequencies (b) versus ferromagnetic interlayer exchange coupling at zero applied field. In panel (b) the weak coupling limit is described by Eq.~\ref{asymSOFTfield1} and Eq.~\ref{asymSOFTfield2}, while the high coupling limit corresponds to Eq.~\ref{asymSOFTfield1largeJ} and Eq.~\ref{asymSOFTfield2largeJ}.
}
\label{figdamping}
\end{figure}

\begin{figure}
\includegraphics[width=9 cm]{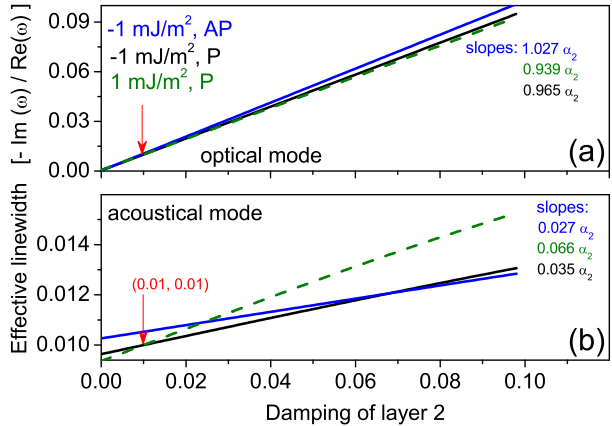}
\caption{Relative Linewidth of the optical (a) and acoustical (b) eigenmode versus the damping parameter of the second layer. The interlayer exchange is varied from ferromagnetic (green curve) to antiferromagnetic (blue curve in the AP remanent state) and black curve in the P state at saturation ($H_z=2.5$~MA/m. The red arrows recall that in the P state the relative linewidth are equal to the damping provided  $\alpha_1=\alpha_2$.}
\label{fig_effectAlpha2}
\end{figure}

However, as soon as the layers are coupled, each eigenmode involves magnetization motion in \textit{both} layers, such that the damping of \textit{both} layers matters to set the linewidth of a given eigenmode \cite{timopheev_dynamic_2014} even at low coupling when the mode can be considered to belong preferentially to one of the layers. 
We have illustrated in Fig.~\ref{figdamping}(a) how the interlayer exchange coupling affects the linewidths of the two modes in a situation with ferromagnetic coupling. Before setting the coupling of coupling, the relative linewidths $\frac{-\Im(\omega_i)}{Re(\omega_i)} $ of the two modes are equal to the $\alpha_i$ as expected. The change of eigenmode relative linewidth is not proportional to the coupling. Tiny couplings ($J << 0.5~\textrm{mJ/m}^2$) do not affect much the relative linewidths that appears to be quadratic with low $J$ [Fig.~\ref{figdamping}(a)]. This contrasts with the eigenmode frequencies [Fig.~\ref{figdamping}(b)] that evolve linearly with $J$ at low coupling (Eq.~\ref{asymSOFTfield1} and Eq.~\ref{asymSOFTfield2}).  
At large coupling the linewidth of the broadest linewidth mode is substantially reduced while the linewidth of the narrowest mode is substantially enlarged [Fig.~\ref{figdamping}(a)]; this continued linewidth broadening of the acoustical mode goes on despite the fact that the frequency of this mode does only evolve in an asymptotic way (see Fig.~\ref{figdamping}(b), green curve).

Let us see to what extent we can manipulate the linewidth of the optical and acoustical eigenmodes when playing with only one damping parameter, for instance $\alpha_2$ (Fig.~\ref{fig_effectAlpha2}). It appears that the relative linewidths are quasi affine functions of the damping parameters when in the collinear P and AP states. In the P state situations, an equality of the damping in the two layers leads a strict equality of the linewidths with the damping for the two modes, as stressed by the red arrows in Fig.~\ref{fig_effectAlpha2}. Comparatively, the AP states have a larger linewidth, larger that the damping of common to the two layers. 

Indeed the impact of the damping factors onto the linewidths depends significantly on the magnetization arrangement (P, AP or NCS) [Fig.~\ref{figLinewidth}(a)]. This is because the amplitudes of dynamical magnetization in each layer depend on their arrangements .
When $\alpha_2>> \alpha_1$ the largest influence of $\alpha_2$ on the acoustical mode linewidth is obtained for ferromagnetic coupling in the P state [green curve in Fig.~\ref{fig_effectAlpha2}(b)]. 
Generally the linewidths of the eigenmodes in the AP states are different from the ones of the P state. \\
Such a configuration-dependent linewidth has been found in the past in various in-plane magnetized systems including Fe/Au/YIG bilayers\cite{heinrich_spin_2011} and all-metallic spin-valves \cite{joyeux_configuration_2011, salikhov_configurational_2012}; in these weakly exchange-coupled systems, the configuration dependence of the linewidth was attributed to a configuration-dependent contribution of spin pumping coupling \cite{kim_magnetization_2005, chiba_magnetization_2015}, but our results suggest that it could also be partly attributed to the interlayer exchange coupling, in line with the conclusions drawn for in-plane magnetized layers in ref.~\onlinecite{timopheev_dynamic_2014}. A way to discriminate between spin-pumping-induced and interlayer-exchange-induced configuration-dependence of the linewidth is that the first one is substantially enhanced when near a crossing of the acoustical and optical eigenexcitation frequencies, while the second one has a broadband impact i.e. it affects the linewidths irrespective of the frequency distance between the two eigenmodes.

Our findings have consequences for the metrology of the interlayer exchange coupling in the ferromagnetic case. In the low coupling limit, the eigenmode frequencies depend linearly on the coupling, such that using the eigenmode frequencies is by far the best way to deduce the coupling strength. The situation is opposite at large couplings where one would like to use the highest frequency mode. Unfortunately in many experiments so far the limited sensitivity renders the detection of the optical mode challenging while the acoustical mode can be characterized in greater detail \cite{devolder_joint_2015}. 
The asymptotic dependence (see Fig.~\ref{figdamping}(b), green curve) of the frequency of the acoustical mode at large coupling makes it ineffective to deduce more than a lower bound for $J$. An alternative method is to benefit from the pronounced change of the eigenmode  linewidth with the coupling at large $J$ : the linewidth of the acoustical mode can be used to quantify the coupling provided the damping parameters are known. 

Besides, our findings have also consequences for the material engineering in STT-MRAM technologies, where stable fixed layers are needed for proper device operation. Here "stable" means both stable against thermal fluctuations -- implying large eigenmode frequency at remanence (or equivalently large coercivity) -- and stable against spin-torque, which requires in addition the largest possible linewidth to prevent auto-oscillation or switching.
The typical situation \cite{sun_effect_2011, gan_perpendicular_2014, worledge_spin_2011, jung_interlayer_2012, devolder_performance_2013, moriyama_tunnel_2010, you_co/ni_2012, ishikawa_magnetic_2013, ishikawa_co/pt_2014} is that layer 1 is an FeCoB-based spin polarizing layer ($i=1$) of an MTJ. This nominally fixed layer needs to be further stabilized, which is usually \cite{sun_effect_2011, gan_perpendicular_2014, worledge_spin_2011, Jung_interlayer_2012, devolder_performance_2013, moriyama_tunnel_2010, you_co/ni_2012, ishikawa_magnetic_2013, ishikawa_co/pt_2014} done by coupling it ferromagnetically through Ta \cite{le_goff_optimization_2015} with harder reference layers ($i=2$) that are usually made of thicker and heavily damped high anisotropy materials, i.e. $\alpha_2 >> \alpha_1$. From Fig.~\ref{figdamping}(b) we see that the thermal stability of the layer 1 can be increased through the coupling, but the zero field frequency of its acoustical excitation is subject to saturation. Fortunately it is possible to obtain a further substantial increase of its stability against spin-torque (i.e. further increase of the effective damping) by increasing further the coupling (see Fig.~\ref{figdamping}(a), green curve). As a result, a larger stabilizing effect can be obtained if the interlayer exchange coupling is further increased despite the fact that the frequency of the acoustical mode converges asymptotically (Eq.~\ref{asymSOFTfield1largeJ}) to a higher limit.

\section{Summary and conclusion}
In the case of strong ferromagnetic coupling between magnetic layers with perpendicular anisotropy, the measurement of the strength of the coupling can not rely on the measurement of the coercivities, as they are usually extrinsic and only weakly depend on the strength of the coupling at large coupling. Instead, we propose to use the frequency of the ferromagnetic resonance modes along the hysteresis loops, because their frequencies are always indicative of the intrinsic properties. One can use the fact that the optical mode has a frequency that is strongly sensitive to the strength of the coupling or the fact that the acoustical mode linewidth can also be largely dependent on the coupling. We have given analytical expressions derived in limit cases to analyze the ferromagnetic resonance frequencies, their linewidth and their meaning for a soft/hard perpendicular magnetized composite. Our results can be used to derive the minimum interlayer exchange coupling needed for the proper pinning of the softest of the two layers, and to optimize their stability against spin transfer torques.

\section*{acknowledgment}
I acknowledge support from the Samsung Global MRAM Innovation program especially V. Nikitin, who provided samples on which we could test and validate the proposed method. Discussions with Joo-Von Kim contributed to clarify some aspects related to eigenmode linewidth.


\begin{thebibliography}{26}%
\makeatletter
\providecommand \@ifxundefined [1]{%
 \@ifx{#1\undefined}
}%
\providecommand \@ifnum [1]{%
 \ifnum #1\expandafter \@firstoftwo
 \else \expandafter \@secondoftwo
 \fi
}%
\providecommand \@ifx [1]{%
 \ifx #1\expandafter \@firstoftwo
 \else \expandafter \@secondoftwo
 \fi
}%
\providecommand \natexlab [1]{#1}%
\providecommand \enquote  [1]{``#1''}%
\providecommand \bibnamefont  [1]{#1}%
\providecommand \bibfnamefont [1]{#1}%
\providecommand \citenamefont [1]{#1}%
\providecommand \href@noop [0]{\@secondoftwo}%
\providecommand \href [0]{\begingroup \@sanitize@url \@href}%
\providecommand \@href[1]{\@@startlink{#1}\@@href}%
\providecommand \@@href[1]{\endgroup#1\@@endlink}%
\providecommand \@sanitize@url [0]{\catcode `\\12\catcode `\$12\catcode
  `\&12\catcode `\#12\catcode `\^12\catcode `\_12\catcode `\%12\relax}%
\providecommand \@@startlink[1]{}%
\providecommand \@@endlink[0]{}%
\providecommand \url  [0]{\begingroup\@sanitize@url \@url }%
\providecommand \@url [1]{\endgroup\@href {#1}{\urlprefix }}%
\providecommand \urlprefix  [0]{URL }%
\providecommand \Eprint [0]{\href }%
\providecommand \doibase [0]{http://dx.doi.org/}%
\providecommand \selectlanguage [0]{\@gobble}%
\providecommand \bibinfo  [0]{\@secondoftwo}%
\providecommand \bibfield  [0]{\@secondoftwo}%
\providecommand \translation [1]{[#1]}%
\providecommand \BibitemOpen [0]{}%
\providecommand \bibitemStop [0]{}%
\providecommand \bibitemNoStop [0]{.\EOS\space}%
\providecommand \EOS [0]{\spacefactor3000\relax}%
\providecommand \BibitemShut  [1]{\csname bibitem#1\endcsname}%
\let\auto@bib@innerbib\@empty
\bibitem [{\citenamefont {Ikeda}\ \emph {et~al.}(2010)\citenamefont {Ikeda},
  \citenamefont {Miura}, \citenamefont {Yamamoto}, \citenamefont {Mizunuma},
  \citenamefont {Gan}, \citenamefont {Endo}, \citenamefont {Kanai},
  \citenamefont {Hayakawa}, \citenamefont {Matsukura},\ and\ \citenamefont
  {Ohno}}]{ikeda_perpendicular-anisotropy_2010}%
  \BibitemOpen
  \bibfield  {author} {\bibinfo {author} {\bibfnamefont {S.}~\bibnamefont
  {Ikeda}}, \bibinfo {author} {\bibfnamefont {K.}~\bibnamefont {Miura}},
  \bibinfo {author} {\bibfnamefont {H.}~\bibnamefont {Yamamoto}}, \bibinfo
  {author} {\bibfnamefont {K.}~\bibnamefont {Mizunuma}}, \bibinfo {author}
  {\bibfnamefont {H.~D.}\ \bibnamefont {Gan}}, \bibinfo {author} {\bibfnamefont
  {M.}~\bibnamefont {Endo}}, \bibinfo {author} {\bibfnamefont {S.}~\bibnamefont
  {Kanai}}, \bibinfo {author} {\bibfnamefont {J.}~\bibnamefont {Hayakawa}},
  \bibinfo {author} {\bibfnamefont {F.}~\bibnamefont {Matsukura}}, \ and\
  \bibinfo {author} {\bibfnamefont {H.}~\bibnamefont {Ohno}},\ }\href {\doibase
  10.1038/nmat2804} {\bibfield  {journal} {\bibinfo  {journal} {Nature
  Materials}\ }\textbf {\bibinfo {volume} {9}},\ \bibinfo {pages} {721}
  (\bibinfo {year} {2010})}\BibitemShut {NoStop}%
\bibitem [{\citenamefont {Khvalkovskiy}\ \emph {et~al.}(2013)\citenamefont
  {Khvalkovskiy}, \citenamefont {Apalkov}, \citenamefont {Watts}, \citenamefont
  {Chepulskii}, \citenamefont {Beach}, \citenamefont {Ong}, \citenamefont
  {Tang}, \citenamefont {Driskill-Smith}, \citenamefont {Butler}, \citenamefont
  {Visscher}, \citenamefont {Lottis}, \citenamefont {Chen}, \citenamefont
  {Nikitin},\ and\ \citenamefont {Krounbi}}]{khvalkovskiy_basic_2013}%
  \BibitemOpen
  \bibfield  {author} {\bibinfo {author} {\bibfnamefont {A.~V.}\ \bibnamefont
  {Khvalkovskiy}}, \bibinfo {author} {\bibfnamefont {D.}~\bibnamefont
  {Apalkov}}, \bibinfo {author} {\bibfnamefont {S.}~\bibnamefont {Watts}},
  \bibinfo {author} {\bibfnamefont {R.}~\bibnamefont {Chepulskii}}, \bibinfo
  {author} {\bibfnamefont {R.~S.}\ \bibnamefont {Beach}}, \bibinfo {author}
  {\bibfnamefont {A.}~\bibnamefont {Ong}}, \bibinfo {author} {\bibfnamefont
  {X.}~\bibnamefont {Tang}}, \bibinfo {author} {\bibfnamefont {A.}~\bibnamefont
  {Driskill-Smith}}, \bibinfo {author} {\bibfnamefont {W.~H.}\ \bibnamefont
  {Butler}}, \bibinfo {author} {\bibfnamefont {P.~B.}\ \bibnamefont
  {Visscher}}, \bibinfo {author} {\bibfnamefont {D.}~\bibnamefont {Lottis}},
  \bibinfo {author} {\bibfnamefont {E.}~\bibnamefont {Chen}}, \bibinfo {author}
  {\bibfnamefont {V.}~\bibnamefont {Nikitin}}, \ and\ \bibinfo {author}
  {\bibfnamefont {M.}~\bibnamefont {Krounbi}},\ }\href {\doibase
  10.1088/0022-3727/46/7/074001} {\bibfield  {journal} {\bibinfo  {journal}
  {Journal of Physics D: Applied Physics}\ }\textbf {\bibinfo {volume} {46}},\
  \bibinfo {pages} {074001} (\bibinfo {year} {2013})}\BibitemShut {NoStop}%
\bibitem [{\citenamefont {Sun}\ \emph {et~al.}(2011)\citenamefont {Sun},
  \citenamefont {Robertazzi}, \citenamefont {Nowak}, \citenamefont
  {Trouilloud}, \citenamefont {Hu}, \citenamefont {Abraham}, \citenamefont
  {Gaidis}, \citenamefont {Brown}, \citenamefont {O~Sullivan}, \citenamefont
  {Gallagher},\ and\ \citenamefont {Worledge}}]{sun_effect_2011}%
  \BibitemOpen
  \bibfield  {author} {\bibinfo {author} {\bibfnamefont {J.}~\bibnamefont
  {Sun}}, \bibinfo {author} {\bibfnamefont {R.}~\bibnamefont {Robertazzi}},
  \bibinfo {author} {\bibfnamefont {J.}~\bibnamefont {Nowak}}, \bibinfo
  {author} {\bibfnamefont {P.}~\bibnamefont {Trouilloud}}, \bibinfo {author}
  {\bibfnamefont {G.}~\bibnamefont {Hu}}, \bibinfo {author} {\bibfnamefont
  {D.}~\bibnamefont {Abraham}}, \bibinfo {author} {\bibfnamefont
  {M.}~\bibnamefont {Gaidis}}, \bibinfo {author} {\bibfnamefont
  {S.}~\bibnamefont {Brown}}, \bibinfo {author} {\bibfnamefont
  {E.}~\bibnamefont {O~Sullivan}}, \bibinfo {author} {\bibfnamefont
  {W.}~\bibnamefont {Gallagher}}, \ and\ \bibinfo {author} {\bibfnamefont
  {D.}~\bibnamefont {Worledge}},\ }\href {\doibase 10.1103/PhysRevB.84.064413}
  {\bibfield  {journal} {\bibinfo  {journal} {Physical Review B}\ }\textbf
  {\bibinfo {volume} {84}},\ \bibinfo {pages} {064413} (\bibinfo {year}
  {2011})}\BibitemShut {NoStop}%
\bibitem [{\citenamefont {Gan}\ \emph {et~al.}(2014)\citenamefont {Gan},
  \citenamefont {Malmhall}, \citenamefont {Wang}, \citenamefont {Yen},
  \citenamefont {Zhang}, \citenamefont {Wang}, \citenamefont {Zhou},
  \citenamefont {Hao}, \citenamefont {Jung}, \citenamefont {Satoh},\ and\
  \citenamefont {Huai}}]{gan_perpendicular_2014}%
  \BibitemOpen
  \bibfield  {author} {\bibinfo {author} {\bibfnamefont {H.}~\bibnamefont
  {Gan}}, \bibinfo {author} {\bibfnamefont {R.}~\bibnamefont {Malmhall}},
  \bibinfo {author} {\bibfnamefont {Z.}~\bibnamefont {Wang}}, \bibinfo {author}
  {\bibfnamefont {B.~K.}\ \bibnamefont {Yen}}, \bibinfo {author} {\bibfnamefont
  {J.}~\bibnamefont {Zhang}}, \bibinfo {author} {\bibfnamefont
  {X.}~\bibnamefont {Wang}}, \bibinfo {author} {\bibfnamefont {Y.}~\bibnamefont
  {Zhou}}, \bibinfo {author} {\bibfnamefont {X.}~\bibnamefont {Hao}}, \bibinfo
  {author} {\bibfnamefont {D.}~\bibnamefont {Jung}}, \bibinfo {author}
  {\bibfnamefont {K.}~\bibnamefont {Satoh}}, \ and\ \bibinfo {author}
  {\bibfnamefont {Y.}~\bibnamefont {Huai}},\ }\href {\doibase
  10.1063/1.4901439} {\bibfield  {journal} {\bibinfo  {journal} {Applied
  Physics Letters}\ }\textbf {\bibinfo {volume} {105}},\ \bibinfo {pages}
  {192403} (\bibinfo {year} {2014})}\BibitemShut {NoStop}%
\bibitem [{\citenamefont {Worledge}\ \emph {et~al.}(2011)\citenamefont
  {Worledge}, \citenamefont {Hu}, \citenamefont {Abraham}, \citenamefont {Sun},
  \citenamefont {Trouilloud}, \citenamefont {Nowak}, \citenamefont {Brown},
  \citenamefont {Gaidis}, \citenamefont {OÕSullivan},\ and\ \citenamefont
  {Robertazzi}}]{worledge_spin_2011}%
  \BibitemOpen
  \bibfield  {author} {\bibinfo {author} {\bibfnamefont {D.~C.}\ \bibnamefont
  {Worledge}}, \bibinfo {author} {\bibfnamefont {G.}~\bibnamefont {Hu}},
  \bibinfo {author} {\bibfnamefont {D.~W.}\ \bibnamefont {Abraham}}, \bibinfo
  {author} {\bibfnamefont {J.~Z.}\ \bibnamefont {Sun}}, \bibinfo {author}
  {\bibfnamefont {P.~L.}\ \bibnamefont {Trouilloud}}, \bibinfo {author}
  {\bibfnamefont {J.}~\bibnamefont {Nowak}}, \bibinfo {author} {\bibfnamefont
  {S.}~\bibnamefont {Brown}}, \bibinfo {author} {\bibfnamefont {M.~C.}\
  \bibnamefont {Gaidis}}, \bibinfo {author} {\bibfnamefont {E.~J.}\
  \bibnamefont {OÕSullivan}}, \ and\ \bibinfo {author} {\bibfnamefont
  {R.~P.}\ \bibnamefont {Robertazzi}},\ }\href {\doibase 10.1063/1.3536482}
  {\bibfield  {journal} {\bibinfo  {journal} {Applied Physics Letters}\
  }\textbf {\bibinfo {volume} {98}},\ \bibinfo {pages} {022501} (\bibinfo
  {year} {2011})}\BibitemShut {NoStop}%
\bibitem [{\citenamefont {Jung}, \citenamefont {Lim},\ and\ \citenamefont
  {Lee}(2012)}]{jung_interlayer_2012}%
  \BibitemOpen
  \bibfield  {author} {\bibinfo {author} {\bibfnamefont {J.~H.}\ \bibnamefont
  {Jung}}, \bibinfo {author} {\bibfnamefont {S.~H.}\ \bibnamefont {Lim}}, \
  and\ \bibinfo {author} {\bibfnamefont {S.~R.}\ \bibnamefont {Lee}},\ }\href
  {\doibase 10.1063/1.4770300} {\bibfield  {journal} {\bibinfo  {journal}
  {Applied Physics Letters}\ }\textbf {\bibinfo {volume} {101}},\ \bibinfo
  {pages} {242403} (\bibinfo {year} {2012})}\BibitemShut {NoStop}%
\bibitem [{\citenamefont {Devolder}\ \emph
  {et~al.}(2013{\natexlab{a}})\citenamefont {Devolder}, \citenamefont {Garcia},
  \citenamefont {Agnus}, \citenamefont {Manfrini}, \citenamefont
  {Cornelissen},\ and\ \citenamefont {Min}}]{devolder_performance_2013}%
  \BibitemOpen
  \bibfield  {author} {\bibinfo {author} {\bibfnamefont {T.}~\bibnamefont
  {Devolder}}, \bibinfo {author} {\bibfnamefont {K.}~\bibnamefont {Garcia}},
  \bibinfo {author} {\bibfnamefont {G.}~\bibnamefont {Agnus}}, \bibinfo
  {author} {\bibfnamefont {M.}~\bibnamefont {Manfrini}}, \bibinfo {author}
  {\bibfnamefont {S.}~\bibnamefont {Cornelissen}}, \ and\ \bibinfo {author}
  {\bibfnamefont {T.}~\bibnamefont {Min}},\ }\href {\doibase 10.1063/1.4826563}
  {\bibfield  {journal} {\bibinfo  {journal} {Applied Physics Letters}\
  }\textbf {\bibinfo {volume} {103}},\ \bibinfo {pages} {182402} (\bibinfo
  {year} {2013}{\natexlab{a}})}\BibitemShut {NoStop}%
\bibitem [{\citenamefont {Moriyama}\ \emph {et~al.}(2010)\citenamefont
  {Moriyama}, \citenamefont {Gudmundsen}, \citenamefont {Huang}, \citenamefont
  {Liu}, \citenamefont {Muller}, \citenamefont {Ralph},\ and\ \citenamefont
  {Buhrman}}]{moriyama_tunnel_2010}%
  \BibitemOpen
  \bibfield  {author} {\bibinfo {author} {\bibfnamefont {T.}~\bibnamefont
  {Moriyama}}, \bibinfo {author} {\bibfnamefont {T.~J.}\ \bibnamefont
  {Gudmundsen}}, \bibinfo {author} {\bibfnamefont {P.~Y.}\ \bibnamefont
  {Huang}}, \bibinfo {author} {\bibfnamefont {L.}~\bibnamefont {Liu}}, \bibinfo
  {author} {\bibfnamefont {D.~A.}\ \bibnamefont {Muller}}, \bibinfo {author}
  {\bibfnamefont {D.~C.}\ \bibnamefont {Ralph}}, \ and\ \bibinfo {author}
  {\bibfnamefont {R.~A.}\ \bibnamefont {Buhrman}},\ }\href {\doibase
  10.1063/1.3481798} {\bibfield  {journal} {\bibinfo  {journal} {Applied
  Physics Letters}\ }\textbf {\bibinfo {volume} {97}},\ \bibinfo {pages}
  {072513} (\bibinfo {year} {2010})}\BibitemShut {NoStop}%
\bibitem [{\citenamefont {You}\ \emph {et~al.}(2012)\citenamefont {You},
  \citenamefont {Sousa}, \citenamefont {Bandiera}, \citenamefont {Rodmacq},\
  and\ \citenamefont {Dieny}}]{you_co/ni_2012}%
  \BibitemOpen
  \bibfield  {author} {\bibinfo {author} {\bibfnamefont {L.}~\bibnamefont
  {You}}, \bibinfo {author} {\bibfnamefont {R.~C.}\ \bibnamefont {Sousa}},
  \bibinfo {author} {\bibfnamefont {S.}~\bibnamefont {Bandiera}}, \bibinfo
  {author} {\bibfnamefont {B.}~\bibnamefont {Rodmacq}}, \ and\ \bibinfo
  {author} {\bibfnamefont {B.}~\bibnamefont {Dieny}},\ }\href {\doibase
  10.1063/1.4704184} {\bibfield  {journal} {\bibinfo  {journal} {Applied
  Physics Letters}\ }\textbf {\bibinfo {volume} {100}},\ \bibinfo {pages}
  {172411} (\bibinfo {year} {2012})}\BibitemShut {NoStop}%
\bibitem [{\citenamefont {Ishikawa}\ \emph {et~al.}(2013)\citenamefont
  {Ishikawa}, \citenamefont {Sato}, \citenamefont {Yamanouchi}, \citenamefont
  {Ikeda}, \citenamefont {Fukami}, \citenamefont {Matsukura},\ and\
  \citenamefont {Ohno}}]{ishikawa_magnetic_2013}%
  \BibitemOpen
  \bibfield  {author} {\bibinfo {author} {\bibfnamefont {S.}~\bibnamefont
  {Ishikawa}}, \bibinfo {author} {\bibfnamefont {H.}~\bibnamefont {Sato}},
  \bibinfo {author} {\bibfnamefont {M.}~\bibnamefont {Yamanouchi}}, \bibinfo
  {author} {\bibfnamefont {S.}~\bibnamefont {Ikeda}}, \bibinfo {author}
  {\bibfnamefont {S.}~\bibnamefont {Fukami}}, \bibinfo {author} {\bibfnamefont
  {F.}~\bibnamefont {Matsukura}}, \ and\ \bibinfo {author} {\bibfnamefont
  {H.}~\bibnamefont {Ohno}},\ }\href {\doibase 10.1063/1.4798499} {\bibfield
  {journal} {\bibinfo  {journal} {Journal of Applied Physics}\ }\textbf
  {\bibinfo {volume} {113}},\ \bibinfo {pages} {17C721} (\bibinfo {year}
  {2013})}\BibitemShut {NoStop}%
\bibitem [{\citenamefont {Ishikawa}\ \emph {et~al.}(2014)\citenamefont
  {Ishikawa}, \citenamefont {Sato}, \citenamefont {Yamanouchi}, \citenamefont
  {Ikeda}, \citenamefont {Fukami}, \citenamefont {Matsukura},\ and\
  \citenamefont {Ohno}}]{ishikawa_co/pt_2014}%
  \BibitemOpen
  \bibfield  {author} {\bibinfo {author} {\bibfnamefont {S.}~\bibnamefont
  {Ishikawa}}, \bibinfo {author} {\bibfnamefont {H.}~\bibnamefont {Sato}},
  \bibinfo {author} {\bibfnamefont {M.}~\bibnamefont {Yamanouchi}}, \bibinfo
  {author} {\bibfnamefont {S.}~\bibnamefont {Ikeda}}, \bibinfo {author}
  {\bibfnamefont {S.}~\bibnamefont {Fukami}}, \bibinfo {author} {\bibfnamefont
  {F.}~\bibnamefont {Matsukura}}, \ and\ \bibinfo {author} {\bibfnamefont
  {H.}~\bibnamefont {Ohno}},\ }\href {\doibase 10.1063/1.4862724} {\bibfield
  {journal} {\bibinfo  {journal} {Journal of Applied Physics}\ }\textbf
  {\bibinfo {volume} {115}},\ \bibinfo {pages} {17C719} (\bibinfo {year}
  {2014})}\BibitemShut {NoStop}%
\bibitem [{\citenamefont {Swerts}\ \emph {et~al.}(2015)\citenamefont {Swerts},
  \citenamefont {Mertens}, \citenamefont {Lin}, \citenamefont {Couet},
  \citenamefont {Tomczak}, \citenamefont {Sankaran}, \citenamefont {Pourtois},
  \citenamefont {Kim}, \citenamefont {Meersschaut}, \citenamefont {Souriau},
  \citenamefont {Radisic}, \citenamefont {Elshocht}, \citenamefont {Kar},\ and\
  \citenamefont {Furnemont}}]{swerts_beol_2015}%
  \BibitemOpen
  \bibfield  {author} {\bibinfo {author} {\bibfnamefont {J.}~\bibnamefont
  {Swerts}}, \bibinfo {author} {\bibfnamefont {S.}~\bibnamefont {Mertens}},
  \bibinfo {author} {\bibfnamefont {T.}~\bibnamefont {Lin}}, \bibinfo {author}
  {\bibfnamefont {S.}~\bibnamefont {Couet}}, \bibinfo {author} {\bibfnamefont
  {Y.}~\bibnamefont {Tomczak}}, \bibinfo {author} {\bibfnamefont
  {K.}~\bibnamefont {Sankaran}}, \bibinfo {author} {\bibfnamefont
  {G.}~\bibnamefont {Pourtois}}, \bibinfo {author} {\bibfnamefont
  {W.}~\bibnamefont {Kim}}, \bibinfo {author} {\bibfnamefont {J.}~\bibnamefont
  {Meersschaut}}, \bibinfo {author} {\bibfnamefont {L.}~\bibnamefont
  {Souriau}}, \bibinfo {author} {\bibfnamefont {D.}~\bibnamefont {Radisic}},
  \bibinfo {author} {\bibfnamefont {S.~V.}\ \bibnamefont {Elshocht}}, \bibinfo
  {author} {\bibfnamefont {G.}~\bibnamefont {Kar}}, \ and\ \bibinfo {author}
  {\bibfnamefont {A.}~\bibnamefont {Furnemont}},\ }\href {\doibase
  10.1063/1.4923420} {\bibfield  {journal} {\bibinfo  {journal} {Applied
  Physics Letters}\ }\textbf {\bibinfo {volume} {106}},\ \bibinfo {pages}
  {262407} (\bibinfo {year} {2015})}\BibitemShut {NoStop}%
\bibitem [{\citenamefont {Devolder}\ \emph {et~al.}(2016)\citenamefont
  {Devolder}, \citenamefont {Kim}, \citenamefont {Garcia-Sanchez},
  \citenamefont {Swerts}, \citenamefont {Kim}, \citenamefont {Couet},
  \citenamefont {Kar},\ and\ \citenamefont
  {Furnemont}}]{devolder_time-resolved_2016}%
  \BibitemOpen
  \bibfield  {author} {\bibinfo {author} {\bibfnamefont {T.}~\bibnamefont
  {Devolder}}, \bibinfo {author} {\bibfnamefont {J.-V.}\ \bibnamefont {Kim}},
  \bibinfo {author} {\bibfnamefont {F.}~\bibnamefont {Garcia-Sanchez}},
  \bibinfo {author} {\bibfnamefont {J.}~\bibnamefont {Swerts}}, \bibinfo
  {author} {\bibfnamefont {W.}~\bibnamefont {Kim}}, \bibinfo {author}
  {\bibfnamefont {S.}~\bibnamefont {Couet}}, \bibinfo {author} {\bibfnamefont
  {G.}~\bibnamefont {Kar}}, \ and\ \bibinfo {author} {\bibfnamefont
  {A.}~\bibnamefont {Furnemont}},\ }\href {\doibase 10.1103/PhysRevB.93.024420}
  {\bibfield  {journal} {\bibinfo  {journal} {Physical Review B}\ }\textbf
  {\bibinfo {volume} {93}},\ \bibinfo {pages} {024420} (\bibinfo {year}
  {2016})}\BibitemShut {NoStop}%
\bibitem [{\citenamefont {Kanai}\ \emph {et~al.}(2014)\citenamefont {Kanai},
  \citenamefont {Tsujikawa}, \citenamefont {Miura}, \citenamefont {Shirai},
  \citenamefont {Matsukura},\ and\ \citenamefont {Ohno}}]{kanai_magnetic_2014}%
  \BibitemOpen
  \bibfield  {author} {\bibinfo {author} {\bibfnamefont {S.}~\bibnamefont
  {Kanai}}, \bibinfo {author} {\bibfnamefont {M.}~\bibnamefont {Tsujikawa}},
  \bibinfo {author} {\bibfnamefont {Y.}~\bibnamefont {Miura}}, \bibinfo
  {author} {\bibfnamefont {M.}~\bibnamefont {Shirai}}, \bibinfo {author}
  {\bibfnamefont {F.}~\bibnamefont {Matsukura}}, \ and\ \bibinfo {author}
  {\bibfnamefont {H.}~\bibnamefont {Ohno}},\ }\href {\doibase
  10.1063/1.4903296} {\bibfield  {journal} {\bibinfo  {journal} {Applied
  Physics Letters}\ }\textbf {\bibinfo {volume} {105}},\ \bibinfo {pages}
  {222409} (\bibinfo {year} {2014})}\BibitemShut {NoStop}%
\bibitem [{\citenamefont {Wi?niowski}\ \emph {et~al.}(2008)\citenamefont
  {Wi?niowski}, \citenamefont {Almeida}, \citenamefont {Cardoso},
  \citenamefont {Barradas},\ and\ \citenamefont
  {Freitas}}]{wisniowski_effect_2008}%
  \BibitemOpen
  \bibfield  {author} {\bibinfo {author} {\bibfnamefont {P.}~\bibnamefont
  {Wi?niowski}}, \bibinfo {author} {\bibfnamefont {J.~M.}\ \bibnamefont
  {Almeida}}, \bibinfo {author} {\bibfnamefont {S.}~\bibnamefont {Cardoso}},
  \bibinfo {author} {\bibfnamefont {N.~P.}\ \bibnamefont {Barradas}}, \ and\
  \bibinfo {author} {\bibfnamefont {P.~P.}\ \bibnamefont {Freitas}},\ }\href
  {\doibase 10.1063/1.2838626} {\bibfield  {journal} {\bibinfo  {journal}
  {Journal of Applied Physics}\ }\textbf {\bibinfo {volume} {103}},\ \bibinfo
  {pages} {07A910} (\bibinfo {year} {2008})}\BibitemShut {NoStop}%
\bibitem [{\citenamefont {Pashaev}\ and\ \citenamefont
  {Mills}(1991)}]{pashaev_ferromagnetic-resonance_1991}%
  \BibitemOpen
  \bibfield  {author} {\bibinfo {author} {\bibfnamefont {K.~M.}\ \bibnamefont
  {Pashaev}}\ and\ \bibinfo {author} {\bibfnamefont {D.~L.}\ \bibnamefont
  {Mills}},\ }\href {\doibase 10.1103/PhysRevB.43.1187} {\bibfield  {journal}
  {\bibinfo  {journal} {Physical Review B}\ }\textbf {\bibinfo {volume} {43}},\
  \bibinfo {pages} {1187} (\bibinfo {year} {1991})}\BibitemShut {NoStop}%
\bibitem [{\citenamefont {Heinrich}\ \emph {et~al.}(1988)\citenamefont
  {Heinrich}, \citenamefont {Purcell}, \citenamefont {Dutcher}, \citenamefont
  {Urquhart}, \citenamefont {Cochran},\ and\ \citenamefont
  {Arrott}}]{heinrich_structural_1988}%
  \BibitemOpen
  \bibfield  {author} {\bibinfo {author} {\bibfnamefont {B.}~\bibnamefont
  {Heinrich}}, \bibinfo {author} {\bibfnamefont {S.~T.}\ \bibnamefont
  {Purcell}}, \bibinfo {author} {\bibfnamefont {J.~R.}\ \bibnamefont
  {Dutcher}}, \bibinfo {author} {\bibfnamefont {K.~B.}\ \bibnamefont
  {Urquhart}}, \bibinfo {author} {\bibfnamefont {J.~F.}\ \bibnamefont
  {Cochran}}, \ and\ \bibinfo {author} {\bibfnamefont {A.~S.}\ \bibnamefont
  {Arrott}},\ }\href {\doibase 10.1103/PhysRevB.38.12879} {\bibfield  {journal}
  {\bibinfo  {journal} {Physical Review B}\ }\textbf {\bibinfo {volume} {38}},\
  \bibinfo {pages} {12879} (\bibinfo {year} {1988})}\BibitemShut {NoStop}%
\bibitem [{\citenamefont {Devolder}\ \emph
  {et~al.}(2013{\natexlab{b}})\citenamefont {Devolder}, \citenamefont {Ducrot},
  \citenamefont {Adam}, \citenamefont {Barisic}, \citenamefont {Vernier},
  \citenamefont {Kim}, \citenamefont {Ockert},\ and\ \citenamefont
  {Ravelosona}}]{devolder_damping_2013}%
  \BibitemOpen
  \bibfield  {author} {\bibinfo {author} {\bibfnamefont {T.}~\bibnamefont
  {Devolder}}, \bibinfo {author} {\bibfnamefont {P.-H.}\ \bibnamefont
  {Ducrot}}, \bibinfo {author} {\bibfnamefont {J.-P.}\ \bibnamefont {Adam}},
  \bibinfo {author} {\bibfnamefont {I.}~\bibnamefont {Barisic}}, \bibinfo
  {author} {\bibfnamefont {N.}~\bibnamefont {Vernier}}, \bibinfo {author}
  {\bibfnamefont {J.-V.}\ \bibnamefont {Kim}}, \bibinfo {author} {\bibfnamefont
  {B.}~\bibnamefont {Ockert}}, \ and\ \bibinfo {author} {\bibfnamefont
  {D.}~\bibnamefont {Ravelosona}},\ }\href {\doibase 10.1063/1.4775684}
  {\bibfield  {journal} {\bibinfo  {journal} {Applied Physics Letters}\
  }\textbf {\bibinfo {volume} {102}},\ \bibinfo {pages} {022407} (\bibinfo
  {year} {2013}{\natexlab{b}})}\BibitemShut {NoStop}%
\bibitem [{\citenamefont {Timopheev}\ \emph {et~al.}(2014)\citenamefont
  {Timopheev}, \citenamefont {Pogorelov}, \citenamefont {Cardoso},
  \citenamefont {Freitas}, \citenamefont {Kakazei},\ and\ \citenamefont
  {Sobolev}}]{timopheev_dynamic_2014}%
  \BibitemOpen
  \bibfield  {author} {\bibinfo {author} {\bibfnamefont {A.~A.}\ \bibnamefont
  {Timopheev}}, \bibinfo {author} {\bibfnamefont {Y.~G.}\ \bibnamefont
  {Pogorelov}}, \bibinfo {author} {\bibfnamefont {S.}~\bibnamefont {Cardoso}},
  \bibinfo {author} {\bibfnamefont {P.~P.}\ \bibnamefont {Freitas}}, \bibinfo
  {author} {\bibfnamefont {G.~N.}\ \bibnamefont {Kakazei}}, \ and\ \bibinfo
  {author} {\bibfnamefont {N.~A.}\ \bibnamefont {Sobolev}},\ }\href {\doibase
  10.1103/PhysRevB.89.144410} {\bibfield  {journal} {\bibinfo  {journal}
  {Physical Review B}\ }\textbf {\bibinfo {volume} {89}},\ \bibinfo {pages}
  {144410} (\bibinfo {year} {2014})}\BibitemShut {NoStop}%
\bibitem [{\citenamefont {Heinrich}\ \emph {et~al.}(2011)\citenamefont
  {Heinrich}, \citenamefont {Burrowes}, \citenamefont {Montoya}, \citenamefont
  {Kardasz}, \citenamefont {Girt}, \citenamefont {Song}, \citenamefont {Sun},\
  and\ \citenamefont {Wu}}]{heinrich_spin_2011}%
  \BibitemOpen
  \bibfield  {author} {\bibinfo {author} {\bibfnamefont {B.}~\bibnamefont
  {Heinrich}}, \bibinfo {author} {\bibfnamefont {C.}~\bibnamefont {Burrowes}},
  \bibinfo {author} {\bibfnamefont {E.}~\bibnamefont {Montoya}}, \bibinfo
  {author} {\bibfnamefont {B.}~\bibnamefont {Kardasz}}, \bibinfo {author}
  {\bibfnamefont {E.}~\bibnamefont {Girt}}, \bibinfo {author} {\bibfnamefont
  {Y.-Y.}\ \bibnamefont {Song}}, \bibinfo {author} {\bibfnamefont
  {Y.}~\bibnamefont {Sun}}, \ and\ \bibinfo {author} {\bibfnamefont
  {M.}~\bibnamefont {Wu}},\ }\href {\doibase 10.1103/PhysRevLett.107.066604}
  {\bibfield  {journal} {\bibinfo  {journal} {Physical Review Letters}\
  }\textbf {\bibinfo {volume} {107}},\ \bibinfo {pages} {066604} (\bibinfo
  {year} {2011})}\BibitemShut {NoStop}%
\bibitem [{\citenamefont {Joyeux}\ \emph {et~al.}(2011)\citenamefont {Joyeux},
  \citenamefont {Devolder}, \citenamefont {Kim}, \citenamefont {Torre},
  \citenamefont {Eimer},\ and\ \citenamefont
  {Chappert}}]{joyeux_configuration_2011}%
  \BibitemOpen
  \bibfield  {author} {\bibinfo {author} {\bibfnamefont {X.}~\bibnamefont
  {Joyeux}}, \bibinfo {author} {\bibfnamefont {T.}~\bibnamefont {Devolder}},
  \bibinfo {author} {\bibfnamefont {J.-V.}\ \bibnamefont {Kim}}, \bibinfo
  {author} {\bibfnamefont {Y.~G. d.~l.}\ \bibnamefont {Torre}}, \bibinfo
  {author} {\bibfnamefont {S.}~\bibnamefont {Eimer}}, \ and\ \bibinfo {author}
  {\bibfnamefont {C.}~\bibnamefont {Chappert}},\ }\href {\doibase
  10.1063/1.3638055} {\bibfield  {journal} {\bibinfo  {journal} {Journal of
  Applied Physics}\ }\textbf {\bibinfo {volume} {110}},\ \bibinfo {pages}
  {063915} (\bibinfo {year} {2011})}\BibitemShut {NoStop}%
\bibitem [{\citenamefont {Salikhov}\ \emph {et~al.}(2012)\citenamefont
  {Salikhov}, \citenamefont {Abrudan}, \citenamefont {BrŸssing}, \citenamefont
  {Gross}, \citenamefont {Luo}, \citenamefont {Westerholt}, \citenamefont
  {Zabel}, \citenamefont {Radu},\ and\ \citenamefont
  {Garifullin}}]{salikhov_configurational_2012}%
  \BibitemOpen
  \bibfield  {author} {\bibinfo {author} {\bibfnamefont {R.}~\bibnamefont
  {Salikhov}}, \bibinfo {author} {\bibfnamefont {R.}~\bibnamefont {Abrudan}},
  \bibinfo {author} {\bibfnamefont {F.}~\bibnamefont {BrŸssing}}, \bibinfo
  {author} {\bibfnamefont {K.}~\bibnamefont {Gross}}, \bibinfo {author}
  {\bibfnamefont {C.}~\bibnamefont {Luo}}, \bibinfo {author} {\bibfnamefont
  {K.}~\bibnamefont {Westerholt}}, \bibinfo {author} {\bibfnamefont
  {H.}~\bibnamefont {Zabel}}, \bibinfo {author} {\bibfnamefont
  {F.}~\bibnamefont {Radu}}, \ and\ \bibinfo {author} {\bibfnamefont {I.~A.}\
  \bibnamefont {Garifullin}},\ }\href {\doibase 10.1103/PhysRevB.86.144422}
  {\bibfield  {journal} {\bibinfo  {journal} {Physical Review B}\ }\textbf
  {\bibinfo {volume} {86}},\ \bibinfo {pages} {144422} (\bibinfo {year}
  {2012})}\BibitemShut {NoStop}%
\bibitem [{\citenamefont {Kim}\ and\ \citenamefont
  {Chappert}(2005)}]{kim_magnetization_2005}%
  \BibitemOpen
  \bibfield  {author} {\bibinfo {author} {\bibfnamefont {J.~V.}\ \bibnamefont
  {Kim}}\ and\ \bibinfo {author} {\bibfnamefont {C.}~\bibnamefont {Chappert}},\
  }\href@noop {} {\bibfield  {journal} {\bibinfo  {journal} {Journal of
  Magnetism and Magnetic Materials}\ }\textbf {\bibinfo {volume} {286}},\
  \bibinfo {pages} {56} (\bibinfo {year} {2005})}\BibitemShut {NoStop}%
\bibitem [{\citenamefont {Chiba}, \citenamefont {Bauer},\ and\ \citenamefont
  {Takahashi}(2015)}]{chiba_magnetization_2015}%
  \BibitemOpen
  \bibfield  {author} {\bibinfo {author} {\bibfnamefont {T.}~\bibnamefont
  {Chiba}}, \bibinfo {author} {\bibfnamefont {G.~E.~W.}\ \bibnamefont {Bauer}},
  \ and\ \bibinfo {author} {\bibfnamefont {S.}~\bibnamefont {Takahashi}},\
  }\href {\doibase 10.1103/PhysRevB.92.054407} {\bibfield  {journal} {\bibinfo
  {journal} {Physical Review B}\ }\textbf {\bibinfo {volume} {92}},\ \bibinfo
  {pages} {054407} (\bibinfo {year} {2015})}\BibitemShut {NoStop}%
\bibitem [{\citenamefont {Devolder}\ \emph {et~al.}(2015)\citenamefont
  {Devolder}, \citenamefont {Goff}, \citenamefont {Eimer},\ and\ \citenamefont
  {Adam}}]{devolder_joint_2015}%
  \BibitemOpen
  \bibfield  {author} {\bibinfo {author} {\bibfnamefont {T.}~\bibnamefont
  {Devolder}}, \bibinfo {author} {\bibfnamefont {A.~L.}\ \bibnamefont {Goff}},
  \bibinfo {author} {\bibfnamefont {S.}~\bibnamefont {Eimer}}, \ and\ \bibinfo
  {author} {\bibfnamefont {J.-P.}\ \bibnamefont {Adam}},\ }\href {\doibase
  10.1063/1.4919089} {\bibfield  {journal} {\bibinfo  {journal} {Journal of
  Applied Physics}\ }\textbf {\bibinfo {volume} {117}},\ \bibinfo {pages}
  {163911} (\bibinfo {year} {2015})}\BibitemShut {NoStop}%
\bibitem [{\citenamefont {Le~Goff}\ \emph {et~al.}(2015)\citenamefont
  {Le~Goff}, \citenamefont {Soucaille}, \citenamefont {Tahmasebi},
  \citenamefont {Swerts}, \citenamefont {Furnemont},\ and\ \citenamefont
  {Devolder}}]{le_goff_optimization_2015}%
  \BibitemOpen
  \bibfield  {author} {\bibinfo {author} {\bibfnamefont {A.}~\bibnamefont
  {Le~Goff}}, \bibinfo {author} {\bibfnamefont {R.}~\bibnamefont {Soucaille}},
  \bibinfo {author} {\bibfnamefont {T.}~\bibnamefont {Tahmasebi}}, \bibinfo
  {author} {\bibfnamefont {J.}~\bibnamefont {Swerts}}, \bibinfo {author}
  {\bibfnamefont {A.}~\bibnamefont {Furnemont}}, \ and\ \bibinfo {author}
  {\bibfnamefont {T.}~\bibnamefont {Devolder}},\ }\href {\doibase
  10.7567/JJAP.54.090302} {\bibfield  {journal} {\bibinfo  {journal} {Japanese
  Journal of Applied Physics}\ }\textbf {\bibinfo {volume} {54}},\ \bibinfo
  {pages} {090302} (\bibinfo {year} {2015})}\BibitemShut {NoStop}%
\end{thebibliography}
%

\end{document}